\documentclass{article}
\usepackage{aas2pp4,tighten}

\def\mathmode#1{\ifmmode{#1}\else{$#1$}\fi}

\newcommand\teff{\mathmode{{T}_{\rm eff}}}
\newcommand\feh{{\rm [Fe/H]}}
\newcommand\etal{{\it et~al.}}

\def\la{\mathrel{\hbox{\rlap{\hbox{\lower4pt\hbox{$\sim$}}}\hbox{$<$}}}}
\def\ga{\mathrel{\hbox{\rlap{\hbox{\lower4pt\hbox{$\sim$}}}\hbox{$>$}}}}

\newcommand\mbfive{\mathmode{m(162)}}
\newcommand\bfivev{\mathmode{(162-V)}}

\def\msun{\mathmode{{\rm M}_\odot}}
\def\flun{\mathmode{ {\rm erg\ s}^{-1}\ {\rm cm}^{-2}\ {\rm
\AA}^{-1}}}
\def\rhalf{\mathmode{{r}_{0.5}}}

\slugcomment{ Submitted to the Astronomical Journal}

\begin{document}

\title{Ultraviolet Imaging of the Globular Cluster 47 Tucanae}

\author{Robert W. O'Connell\altaffilmark{1}, Ben Dorman\altaffilmark{1,2},
Ronak Y. Shah\altaffilmark{1}, Robert T. Rood\altaffilmark{1},}
\author{Wayne B. Landsman\altaffilmark{3}, Ralph C.
Bohlin\altaffilmark{4}, Susan G. Neff\altaffilmark{2},}
\author{Morton S. Roberts\altaffilmark{5}, Andrew M. Smith\altaffilmark{2},
and Theodore P. Stecher\altaffilmark{2}}

\altaffiltext{1}{Astronomy Dept, University of Virginia,
	P.O.Box 3818, Charlottesville, VA 22903-0818.  Electronic mail: rwo@virginia.edu;
dorman@parfait.gsfc.nasa.gov; rys3p@virginia.edu; rtr@virginia.edu}
\altaffiltext{2}{Laboratory for Astronomy \& Solar Physics, 
Code 681, NASA/GSFC, Greenbelt MD 20771.  Electronic mail: neff@uit.gsfc.nasa.gov;
asmith@uit.gsfc.nasa.gov; stecher@uit.gsfc.nasa.gov}
\altaffiltext{3}{Hughes STX Corporation, 
Code 681, NASA/GSFC, Greenbelt MD 20771; Electronic mail: landsman@uit.gsfc.nasa.gov}
\altaffiltext{4}{Space Telescope Science Institute, 3700 San Martin Drive,
Baltimore, MD 21218; Electronic mail: bohlin@stsci.edu}
\altaffiltext{5}{National Radio Astronomy Observatory, 
Charlottesville, VA 22903; Electronic mail: mroberts@nrao.edu}

\begin{abstract}

We have used the Ultraviolet Imaging Telescope to obtain deep far-UV
(1620 \AA), 40\arcmin\ diameter images of the prototypical metal-rich
globular cluster 47 Tucanae.  We find a population of about 20 hot
($\teff > 9000$ K) objects near or above the predicted UV luminosity
of the hot horizontal branch (HB) and lying within two half-light
radii of the cluster center.  We believe these are normal hot HB or
post-HB objects rather than interacting binaries or blue stragglers.
IUE spectra of two are consistent with post-HB phases.  These
observations, and recent HST photometry of two other metal-rich
clusters, demonstrate that populations with rich, cool HB's can
nonetheless produce hot HB and post-HB stars.  The cluster center also
contains an unusual diffuse far-UV source which is more extended than
its $V$-band light.  It is possible that this is associated with an
intracluster medium, for which there was earlier infrared and X-ray
evidence, and is produced by C IV emission or scattered light from
grains.

\end{abstract}


\twocolumn

\section{Introduction}

In this paper we report the first wide-field, vacuum ultraviolet
imaging of the bright globular cluster 47 Tucanae, in which we have
detected both a number of UV-bright, hot stars and an unresolved,
extended UV source.  Our observations were made as part of the
Ultraviolet Imaging Telescope (UIT) survey of globular clusters during
the {\it Astro-2} Spacelab mission in March 1995.  The primary
astrophysical motivation for the survey was to understand the
mechanisms which govern the production of hot stars in horizontal
branch (HB) and more advanced evolutionary phases in globular clusters
and their relationship to the ``UV-upturn'' phenomenon in elliptical
galaxies (Burstein et al.~1988, Greggio \& Renzini 1990).  

47 Tuc is the prototype of the class of metal-rich globulars.  It has
[Fe/H]$\,\sim -0.7$ and one of the most reliable absolute age
determinations ($\sim\,$13.5 Gyr; \cite{hhvas}).  It is regularly used
as a template population to compare to other clusters and the upper
giant branches of galaxy color-magnitude diagrams.  At
optical/infrared wavelengths, the integrated spectral energy
distributions of clusters like 47 Tuc are intermediate between those of
the more metal-poor globulars and elliptical galaxies.  However, they
exhibit a unique behavior in the far-UV ($\lambda < 2000$\AA) because
they have the smallest known ratios of far-UV to optical light of any
old populations yet studied (van Albada, de Boer, \& Dickens 1981;
Rich, Minitti, \& Liebert 1993; Dorman, O'Connell, \& Rood 1995).  The
metal-rich globulars therefore represent an important extreme which is
important to understand in trying to identify the underlying mechanism
of the UV-upturn phenomenon in galaxies.  More discussion of these
issues can be found in our companion paper on UV imaging of the
cluster NGC 362 (Dorman et al.~1997) and in Greggio \& Renzini (1990)
and Dorman et al.~(1995).  

Metal-rich globulars are well known for having predominantly red
horizontal branches.  On the optical-band CMD of Hesser et al.~(1987),
47 Tuc has an exclusively red HB, with stars concentrated at $B-V \ga
0.5$.  The bluest HB star is a lone RR Lyr\ae\ star with $\langle B-V
\rangle = 0.40$ ($\teff \sim 6200\,$K) (\cite{csw}).  Recent
optical-band photometry by Montgomery \& Janes (1994, hereafter MJ94)
includes several blue HB candidates in the outer parts of the
cluster.  The only previously-known luminous hot star in 47 Tuc is the
very bright ($V = 10.73$, spectral type B8 III) post-asymptotic giant
branch (PAGB) star labeled ``BS'' by \cite{le74}.

Earlier vacuum-UV studies of 47 Tuc were made by OAO (Welch \& Code
1980), ANS (van Albada et al.~1981), IUE (Rich et al.~1993), and
HST/FOC (e.g.~Paresce et al.~1991; De Marchi, Paresce, \& Ferraro
1993; Paresce, De Marchi, \& Jedrzejewski 1995).  Except for the early
OAO and ANS photometry all these observations were confined to the
inner 20\arcsec\ of the cluster's core (excluding the BS star).
HST/FOC identified a number of blue stragglers (BSS) with $\teff <
9000\,$K, white dwarfs, and several faint interacting binaries, but no
new hot sources near or above HB luminosity.  The IUE spectra revealed
what appeared to be a warm but spatially extended component which was
too faint to consist of objects as bright as the HB.  

\section{Observations and Data Reduction}

A description of the UIT instrument and standard data reduction and
calibration procedures is given in Stecher et al.~(1997) and
\cite{dor97}.  The field of view is 40\arcmin\ diameter.  Observations
were made with a CsI photocathode, which has excellent long-wavelength
(``red leak'') rejection.  We used the B5 filter, which has a peak
wavelength of 1620 \AA\ and a bandwidth of 230 \AA.  In this paper, we
use mainly the longest of several 47 Tuc exposures (frame FUV2708,
exposure 1680.5 sec), which was also the deepest recorded of any
cluster.  The images have been digitized to a scale of
$1\farcs 14$/pixel.  Point sources have FWHM $\sim$ 4\arcsec\ owing to
jitter in the pointing system aboard the Shuttle. Astrometry was
obtained using a combination of the
\cite{tuc92} and \cite{mj94} data for objects in common with the HST
Guide Star Catalog.  We used an Interactive Data Language (IDL)
implementation of DAOPHOT I (\cite{daop}), which has been modified to
accommodate the noise characteristics of film, to derive stellar
photometry.  Typical errors for the photometry are 0.15 mag, including
an uncertainty in the aperture correction of 0.10 mag owing to a variable
PSF.  No correction is necessary for red leaks.  We
quote FUV magnitudes on the monochromatic system, where
$m_{\lambda}(\lambda) =
 -2.5 \log(F_{\lambda}) -21.1$, and $F_{\lambda}$ is in units of \flun.
We refer to B5 magnitudes below as \mbfive\ and $B5-V$ colors as
\bfivev.  Due to the failure of UIT's mid-UV camera on the {\it
Astro-2} mission, however, we were unable to obtain UV colors for our
targets.  Kent Montgomery (Ph.D.~thesis, 1994, Boston University) 
\& Kenneth Janes have kindly provided us with pre-publication CCD BVI
photometry of 47 Tuc, a summary of which is given in MJ94.  
  
As basic parameters for 47 Tuc, we adopt the following (\cite{sw86};
\cite{tkd95}; \cite{sgd93}):  $(m-M)_0 = 13.3$; $E(B-V) = 0.04$;
$(m-M)_{\rm FUV} = 13.62$; $\rhalf = 174\arcsec$ is the half-light
radius in the optical band; and integrated $M_V = -9.4.$ The
optical-band photocenter (J2000) is at $\rm \alpha = 00^h\, 24^m \,
05^s.2,$ $\rm \delta = -72^\circ \, 04\arcmin
\, 51\arcsec.$ In computing the FUV apparent distance modulus, we have
adopted the Galactic UV reddening law of Cardelli, Clayton \& Mathis
(1989) according to which $\rm A(FUV)/E(B-V) = 8.06.$

\section{Hot Stars in 47 Tuc}

The center of our far-UV image of 47 Tuc is shown in Figure 1 (Plate
XX). The brightest object in the field is the previously-known PAGB
star, about 45\arcsec\ SW of the cluster center.  We detect a number
of fainter UV point sources, concentrated within the cluster's
half-light radius (see inset).  Because of the complete suppression of
the cool main sequence and red giant branch stars by the UIT detector
system, we are able to resolve the hot sources in the cluster
throughout the core, which would not be possible at visible
wavelengths with our FWHM.  Surrounding the cluster center (marked)
is a diffuse UV source; this is not a common feature among the
clusters we have imaged with UIT, and we return to it in \S 4.

47 Tuc lies near the Small Magellanic Cloud in projection, and some of
the UV-bright stars lying beyond several half-light radii may be 
SMC main sequence stars.  Our large field of view provides a good
upper limit to the surface density of such contaminants.  

We identified a total of 51 stars in the field with \mbfive\
magnitudes between 9 and 17.  At optical wavelengths with ground-based
resolution, the cluster center is too dense to make
cross-identifications with our UV detections.  However, at larger
radii ($r \ga 110\arcsec$), we have made seven identifications with
blue stars ($B-V < 0.2$) in other data sets, as listed in Table 1.
Three of these (MJ 280, 33410, 38529) are high probability members of
47 Tuc from proper motions (Tucholke 1992); another (Tucholke \# 2497)
is a possible member. Four other objects have MJ94 identifications (MJ
8279, 19945, 25308, 38298).  The
\bfivev\ colors of all these objects are $ < -1.0$, implying $\teff >
10000\,$K, and most are $< -2$.  MJ 38298, with $V = 18.75$ has
$\bfivev = -4$, indicating $\teff \ga 30000\,$K.  Except for MJ 25308
the stars do not have colors and brightnesses consistent with the SMC
main sequence.  A final identification is with the field star HD 2041
at $10\farcm 35$ radius; this is an F6 IV/V star with $\bfivev \sim
+5.8$.  

Positions and photometry for the UIT identifications are
given in Table 1.  Stars in the table are listed in order of distance from
the cluster center.  The first column gives the UIT identification number;
$R$ is the distance in arcseconds from the cluster photocenter; $\Delta X$
is the east-west offset in arcseconds from the cluster center (west
being positive); $\Delta Y$ is the north-south offset in arcseconds from
the cluster center (north being positive); \mbfive\ is the monochromatic
UV magnitude in the B5 filter (1620 \AA\ centroid); $\sigma (162)$ is
the one-sigma uncertainty in the UV magnitude; $Q$ is a subjective
estimate of the reality of the sources, ranging from 4 (certain)
to 1 (marginal); $V$ and $(B - V)$ are the optical magnitudes and colors
from the cross-identifications listed in the Notes column.  In the Notes
column, MJ $=$ Montgomery \& Janes and  T $=$ Tucholke.

Since we do not have colors for most sources, our analysis of the
resolved population is based mainly on the UV luminosity function,
which is shown in Figure 2.  In order to assess the SMC contamination,
we have plotted the luminosity function in surface density units and for two
separate regions of the field:  $r < 2\,\rhalf$, within which the
cluster dominates the counts, and $r > 3\,\rhalf$, where SMC stars may
be important.  18 of the 51 UV sample stars are within the half-light
radius, 22 are within $2\,\rhalf$, and 21 are outside $3\,\rhalf$.
Only 2 of the MJ94 cross-identifications are within $2\,\rhalf$.  The
brightest object is BS, with $\mbfive = 9.98\pm$0.05.  (BS was
observed by Dixon et al.~1995 with the Hopkins Ultraviolet Telescope
spectrometer during the {\it Astro-2} mission, yielding $\teff
\sim 10,500$ K and $\log g < 2.5$.)

\onecolumn

\begin{deluxetable}{crrrrrrrrl}
\tablewidth{0pc}
\tablecaption{Far-UV Photometry for 47 Tucan\ae}
\tablehead{
\colhead{UIT ID}      & \colhead{R ($\arcsec$)} & \colhead{$\Delta X$} &
\colhead{$\Delta Y$} &  \colhead{$m(162)$} &
\colhead{$\sigma (162)$}  &\colhead{Q} & \colhead{$V$} & \colhead{$(B-V)$} & \colhead{Notes}
}
\startdata

  1 &     17.45 &      -6.92 &      16.02 &      15.46 &       0.13 &    4 &  \nodata  &   \nodata  &   \nodata  \nl 
  2 &     22.23 &      18.72 &     -11.99 &      15.58 &       0.11 &    3 &  \nodata  &   \nodata  &   \nodata  \nl 
  3 &     22.47 &      -0.74 &     -22.46 &      14.85 &       0.11 &    4 &  \nodata  &   \nodata  &   \nodata  \nl 
  4 &     38.12 &      33.36 &      18.45 &      16.24 &       0.12 &    2 &  \nodata  &   \nodata  &   \nodata  \nl 
  5 &     45.63 &      43.22 &      14.63 &      17.45 &       0.18 &    2 &  \nodata  &   \nodata  &   \nodata  \nl 
  6 &     50.12 &     -11.14 &     -48.86 &      15.60 &       0.12 &    4 &  \nodata  &   \nodata  &   \nodata  \nl 
  7 &     51.04 &      33.16 &     -38.80 &      10.30 &       0.11 &    4 &   10.73 &     \nodata  &   BS, B8III Post-AGB \nl
  8 &     72.33 &     -72.33 &       0.76 &      16.25 &       0.12 &    3 &  \nodata  &   \nodata  &   \nodata  \nl 
  9 &     80.90 &     -26.53 &      76.43 &      17.02 &       0.13 &    2 &  \nodata  &   \nodata  &   \nodata  \nl 
 10 &     89.15 &     -45.24 &      76.82 &      16.74 &       0.13 &    2 &  \nodata  &   \nodata  &   \nodata  \nl 
 11 &     90.36 &     -71.27 &      55.54 &      17.04 &       0.15 &    2 &  \nodata  &   \nodata  &   \nodata  \nl 
 12 &     94.32 &      13.89 &      93.29 &      16.26 &       0.13 &    2 &  \nodata  &   \nodata  &   \nodata  \nl 
 13 &     98.13 &      24.23 &      95.09 &      16.14 &       0.13 &    2 &  \nodata  &   \nodata  &   \nodata  \nl 
 14 &    104.77 &      89.81 &     -53.95 &      13.62 &       0.10 &    4 &  \nodata  &   \nodata  &   IUE spectrum  \nl 
 15 &    111.66 &      97.00 &     -55.31 &      13.35 &       0.10 &    4 &  15.56 &   $-0.09$  &   MJ19945; IUE spectrum  \nl 
 16 &    131.02 &     -35.70 &     126.07 &      16.99 &       0.14 &    2 &  \nodata  &   \nodata  &   \nodata  \nl 
 17 &    149.33 &    -141.58 &      47.49 &      14.12 &       0.11 &    4 &  \nodata  &   \nodata  &   \nodata  \nl 
 18 &    151.28 &     -22.38 &     149.61 &      16.15 &       0.11 &    3 &  \nodata  &   \nodata  &   \nodata  \nl 
 19 &    178.54 &    -101.29 &     147.03 &      15.15 &       0.11 &    4 &  \nodata  &   \nodata  &   \nodata  \nl 
 20 &    201.63 &     157.03 &     126.48 &      13.56 &       0.10 &    4 &  \nodata  &   \nodata  &   \nodata  \nl 
 21 &    243.60 &    -239.86 &     -42.53 &      16.25 &       0.11 &    3 &  \nodata  &   \nodata  &   \nodata  \nl 
 22 &    336.08 &    -182.41 &    -282.26 &      13.70 &       0.10 &    4 &   14.43 &     $-0.09$ &   MJ33410, T1948  \nl 
 23 &    397.84 &      12.43 &    -397.64 &      16.57 &       0.12 &    4 &  \nodata  &   \nodata  &   \nodata  \nl 
 24 &    411.72 &    -392.66 &    -123.81 &      15.07 &       0.11 &    4 &      18.75 &    \nodata &  MJ38298  \nl 
 25 &    415.33 &     127.51 &    -395.27 &      15.71 &       0.11 &    4 &  \nodata  &   \nodata  &   \nodata  \nl 
 26 &    418.80 &       5.97 &     418.75 &      13.60 &       0.10 &    4 &  \nodata  &   \nodata  &   \nodata  \nl 
 27 &    423.91 &    -422.17 &      38.37 &      14.38 &       0.11 &    4 &  \nodata  &   \nodata  &   \nodata  \nl 
 28 &    432.59 &     409.20 &    -140.31 &      15.56 &       0.11 &    4 &  \nodata  &   \nodata  &   \nodata  \nl 
 29 &    492.10 &    -182.96 &     456.82 &      15.51 &       0.11 &    4 &  \nodata  &   \nodata  &   \nodata  \nl 
 30 &    518.75 &     314.21 &     412.76 &      15.86 &       0.11 &    4 &    18.64 &     $-0.17$ &   MJ8279 \nl 
 31 &    539.12 &    -406.38 &    -354.27 &      13.82 &       0.11 &    4 &  \nodata  &   \nodata  &   \nodata  \nl 
 32 &    558.12 &    -532.38 &    -167.52 &      16.10 &       0.12 &    4 &  \nodata  &   \nodata  &   \nodata  \nl 
 33 &    596.94 &    -161.79 &    -574.60 &      15.79 &       0.23 &    1 &  \nodata  &   \nodata  &   \nodata  \nl 
 34 &    600.95 &    -598.84 &      50.26 &      15.06 &       0.11 &    3 &  \nodata  &   \nodata  &   \nodata  \nl 
 35 &    620.06 &     612.76 &      94.89 &      15.99 &       0.11 &    4 &  \nodata  &   \nodata  &   \nodata  \nl 
 36 &    621.88 &    -476.36 &     399.78 &      13.21 &       0.10 &    4 & 15.01 &     $-0.07$ &    T2497  \nl  
 37 &    622.52 &      16.40 &    -622.31 &      14.99 &       0.11 &    4 &  \nodata  &   \nodata  &   \nodata  \nl 
 38 &    641.20 &     215.96 &    -603.74 &      16.23 &       0.31 &    1 &  \nodata  &   \nodata  &   \nodata  \nl 
 39 &    653.82 &     426.76 &    -495.34 &      16.21 &       0.21 &    1 &  \nodata  &   \nodata  &   \nodata  \nl 
 40 &    673.56 &    -480.69 &     471.83 &      15.70 &       0.11 &    4 &  \nodata  &   \nodata  &   \nodata  \nl 
 41 &    681.51 &     681.49 &       5.08 &      15.93 &       0.22 &    2 &  \nodata  &   \nodata  &   \nodata  \nl 
 42 &    705.00 &    -315.04 &    -630.70 &      16.31 &       0.26 &    1 &  \nodata  &   \nodata  &   \nodata  \nl 
 43 &    789.04 &    -709.61 &    -345.01 &      16.56 &       0.25 &    1 &  \nodata  &   \nodata  &   \nodata  \nl 
 44 &    817.30 &    -811.76 &      94.97 &      15.90 &       0.11 &    3 &  \nodata  &   \nodata  &   \nodata  \nl 
 45 &    821.41 &     721.75 &     392.16 &      13.18 &       0.11 &    4 &  \nodata  &   \nodata  &   \nodata  \nl 
 46 &    864.98 &      -5.77 &    -864.96 &      16.32 &       0.24 &    1 &   15.23 &     $-0.16$ &   MJ280, T300  \nl 
 47 &    906.97 &    -854.97 &    -302.71 &      15.45 &       0.21 &    1 &  \nodata  &   \nodata  &   \nodata  \nl 
 48 &    943.11 &     602.48 &    -725.58 &      14.28 &       0.10 &    4 &  \nodata  &   \nodata  &   \nodata  \nl 
 49 &    947.46 &    -628.16 &    -709.28 &      15.63 &       0.11 &    3 &  \nodata  &   \nodata  &   \nodata  \nl 
 50 &    960.44 &    -414.76 &     866.27 &      14.32 &       0.11 &    3 &  \nodata  &   \nodata  &   \nodata  \nl 
 51 &    962.44 &     156.68 &    -949.60 &      16.18 &       0.11 &    4 &  \nodata  &   \nodata  &   \nodata  \nl 

\enddata
\end{deluxetable}

\twocolumn

\begin{figure}
\setcounter{figure}{1}
\plotone{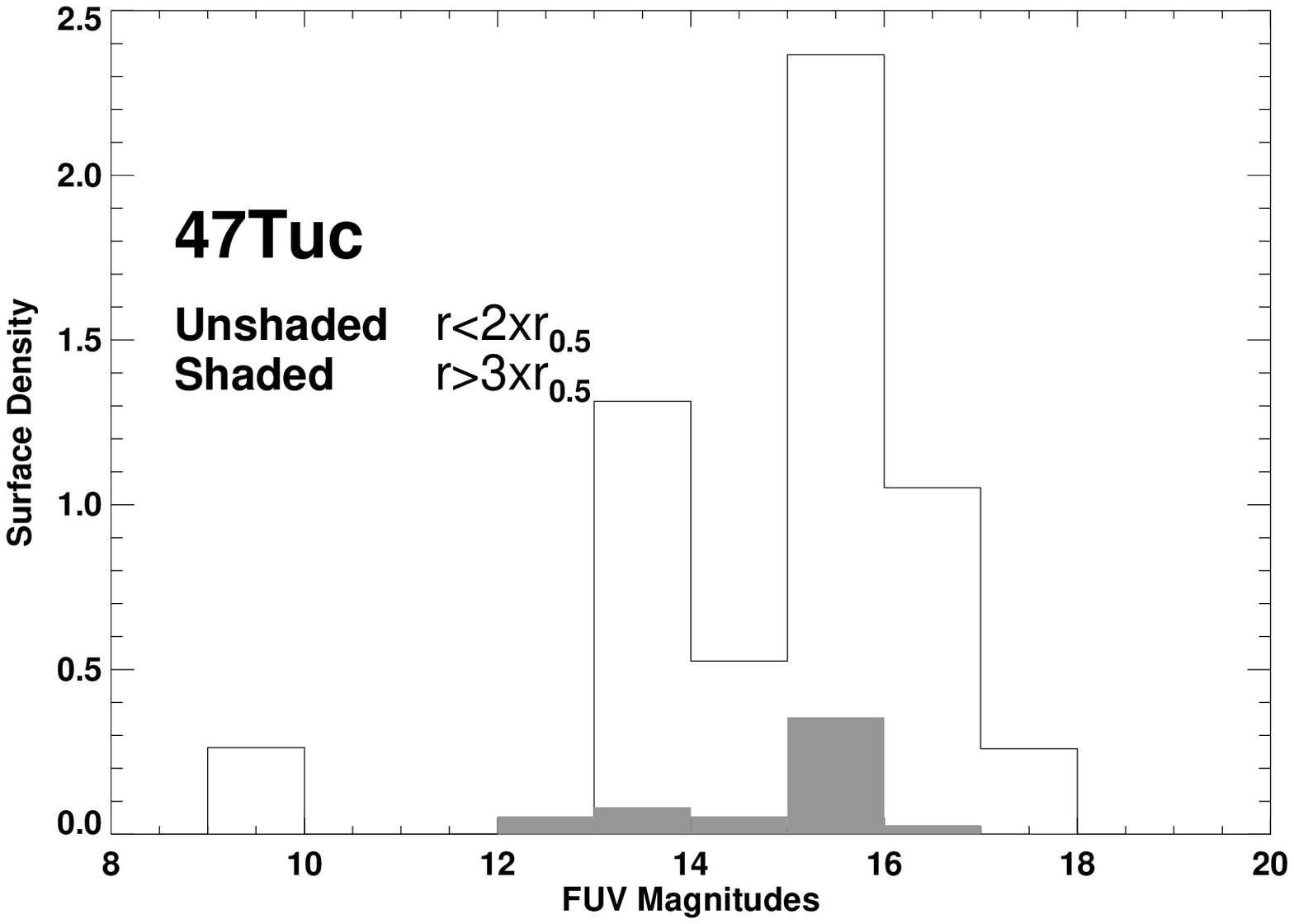}
\caption{ Histogram of the surface densities of UV
identifications for the inner and outer regions of 47 Tuc as a
function of UV brightness.  Units are $10^{-5}\,$ stars per square
arcsecond.  }
\end{figure}

The modes of both the inner and outer histograms plotted occur at
$\mbfive =\,$ 15--16.  It is not clear whether the decline at fainter
magnitudes is intrinsic or is instead produced by the sensitivity
limit of the data.  There is a clear excess (a factor of 7$\times$ at
the mode) in the surface density of detections for the inner field.
Most of the detections here are therefore members of 47 Tuc.  The
astrometric study of Tucholke (1992) found an ``almost a complete
lack of SMC stars'' brighter than a limit of $B
\sim 17$, which is consistent with these results.

The brightest objects other than BS lie at $\mbfive =\,$13--14, which
(see below) is above the expected brightness of the HB in 47 Tuc.  Two
of these supra-HB stars are located within $7\arcsec$ of each other,
approximately $2\arcmin$ southwest of the cluster center.  One of
these (UIT-15) is identified with MJ 19945 ($V = 15.56,\, B-V =
-0.09$) while the other (UIT-14) is unidentified in the optical.  On
11-Sep-1995, we obtained a 400 minute low-dispersion IUE image (SWP
55910), with the large ($10\times 20\arcsec$) aperture centered
between the two stars.  A spectrum for each star was extracted from
the line-by-line IUE image, and is presented in Figure 3.  The UV
spectrum and V magnitude of MJ 19945 are well-fit by a Kurucz (1993)
model with $\teff = 14000\,$K, $\log g = 3.5$, and [Fe/H] $= -0.5$.
The best-fit model of UIT-14 has $\teff = 50000\,$K and $\log g =
5.0$, though the fit is not as good, and the IUE spectrum does not
provide good discrimination for $\teff > 30000\,$K.  UIT-14 is almost
certainly a member of 47 Tuc, given its very high temperature
(characteristic of an sdO star) and proximity to the cluster center.
In both cases, the IUE data confirm the UIT flux measurements and the
expectation that mainly high temperature objects will be present on
the UIT image.  

\begin{figure}
\plotone{47tucfig3.ps}
\caption{ IUE low-resolution, far-UV spectra of
two of the supra-HB identifications in 47 Tuc.  Overplotted are the
best fitting model atmospheres, as described in the text.  UIT-15, the
cooler object, is MJ 19945. }
\end{figure}

For comparison to the luminosity function in Fig. 2, we have created a
synthetic UV-optical CMD for 47 Tuc in observed coordinates (Figure
4).  We transformed the model ZAHB and main sequence for $\feh =-0.78$
from D'Cruz et al.~(1996) by interpolation in the  model
atmosphere grid of Kurucz (1993).  Extinction effects were incorporated
based on the assumptions stated in Sec.~2.  We
have included in the figure the brightest blue stragglers from
the HST/FOC UV photometry of De Marchi et al.~(1993) and also from the
recent HST/WFPC2 optical-band photometry of Sosin et al.~(1997).  In
both cases, we have transformed optical-band photometry to the UV
using the Kurucz atmospheres.  This is relatively crude but suffices
to indicate where the known BSS stars should appear in our counts.
Note that the field covered by the WFPC2 sample extends to $r \leq
110\arcsec$ and includes only 6\% of the area of the $2\,\rhalf$ 
UIT sample.  

\begin{figure}
\plotone{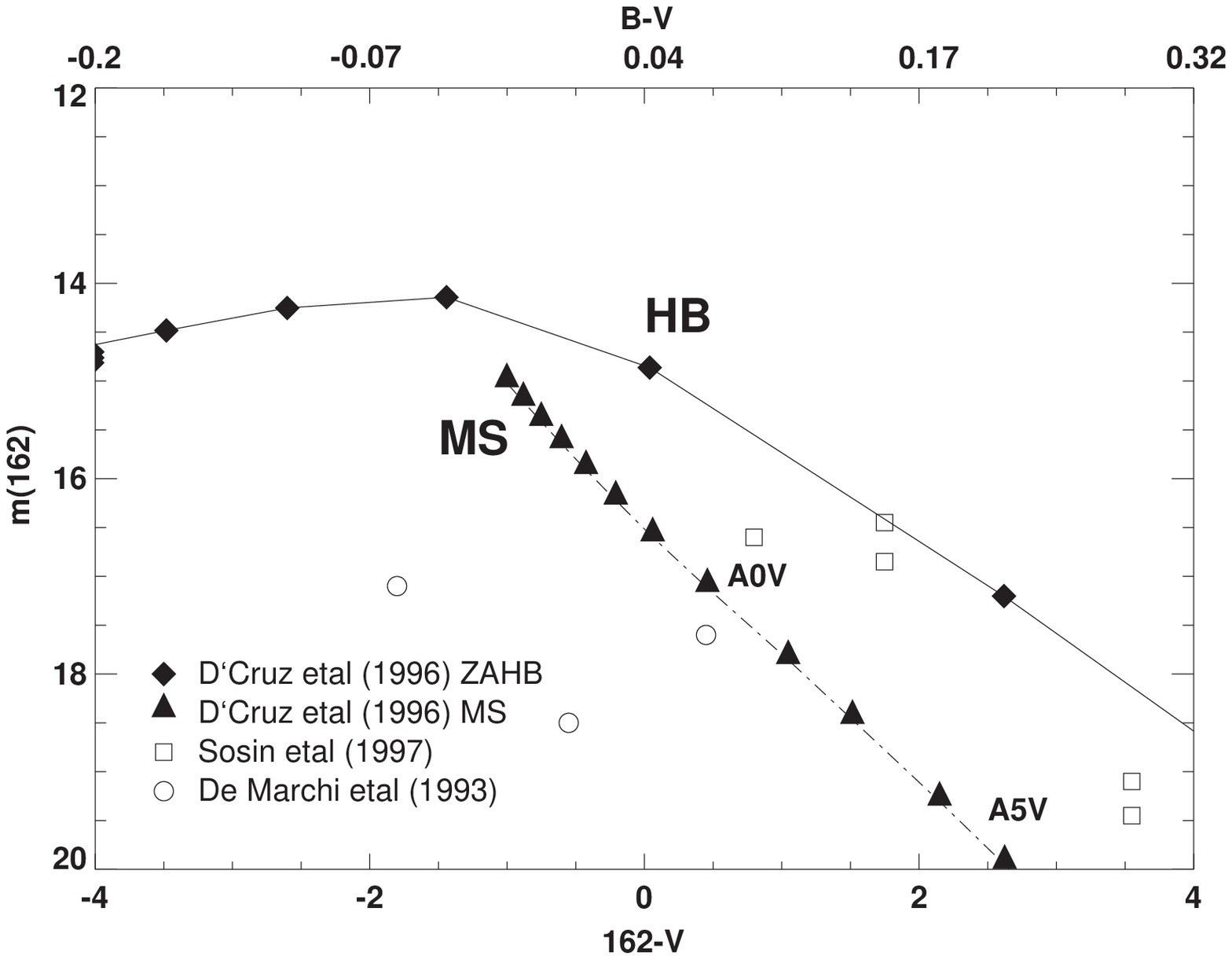}
\caption{ A synthetic UV-optical color-magnitude
diagram for 47 Tuc, in which the theoretical locations of the ZAHB and
the main sequence have been transformed to observational quantities
after adopting standard parameters for the cluster given in the text.
Approximate locations of the brighter blue stragglers detected in HST
observations of the cluster core are plotted individually.  }
\end{figure}

Given our limiting magnitude of $\mbfive \sim 17$, we see from Fig.~4
that we should not detect HB stars cooler than $B-V \sim 0.18$
($\teff \sim 8300\,$K) or blue straggler stars cooler than
$B-V \sim 0.05$ ($\teff \sim 10000\,$K).  The known HST blue stragglers
should all fall at $\mbfive \ga 16.5$.  The great majority of the BSS
which Paresce et al.~(1991) detected in the core of 47 Tuc with a
surface density of $\sim 0.05\, {\rm arcsec}^{-2}$ are too cool to be
detected here, but some of the stars in the two faintest bins of
Fig.~2 are probably BSS.  

We cannot make an unambiguous identification of the kind of stars in
the most populated bin ($m[162] =\,$ 15--16) in our luminosity
function (Fig.~2).  A well populated extreme HB branch (with $B-V <
-0.05$) should appear about one magnitude brighter (see Fig.~4), and
the two independent HST studies agree that the BSS objects probably
fall at least one magnitude fainter.  Based on the calibration in Fig.
4, the modal bin corresponds to HB stars in the range $B-V \sim\,$
0.0--0.1, $V > 16$.  It is not clear how well this region has been
explored in existing optical-band CMD's.

It is possible that the mode of the luminosity function in Fig.~2 does
correspond to the extreme horizontal branch if the theoretical
calibration in Fig.~4 slightly overpredicts the hot HB luminosity.  We
encountered this circumstance in several other clusters observed with
UIT (e.g.~NGC 1851, Parise et al.~1994; and $\omega$ Cen, Whitney et
al.~1994).  The number of probable post-HB stars in the luminosity
function also favors this interpretation.  The models in Fig.~4
predict that extreme HB stars should not be brighter than $\mbfive
\sim 14$.  The five objects in the $\mbfive =\,$ 13--14 bin are
therefore supra-HB stars, probably in post-HB evolutionary phases
moving coolward in the CMD toward the asymptotic giant branch or
upward in the CMD as ``AGB-Manqu\`e'' objects (e.g.\ Greggio \& Renzini
1990; Dorman, Rood, \& O'Connell 1993).  The presence of a few objects
of this type has also been inferred in other metal-rich clusters from
IUE spectroscopy (Rich et al.~1993), and the RR Lyrae object in 47 Tuc
is probably another example (Carney et al.~1993).  The evolutionary
lifetimes of these post-HB phases, however, are such that one expects
about 4 times as many blue HB precursors.  Only by interpreting the
mode of the luminosity function in Fig. 2 as representing the blue HB
can the density of supra-HB stars be made roughly consistent with
canonical evolutionary theory. 

An alternative intepretation of the supra-HB stars as interacting
binaries (e.g. Rich et al.~1993) is unlikely, given the facts that
almost all such objects have luminosities below the HB and that UIT-14
and MJ 19445 seem to have normal spectra for post-HB phases.  No
emission lines are present in the spectra.  A more exotic
interpretation of the supra-HB stars is that they are due to binary
mergers of helium white dwarfs (Iben 1990, Bailyn 1995), which is a
process that can yield core helium-burning stars with core masses of
up to $0.8\msun$.  This is the preferred explanation for two hot
supra-HB stars in M3, which lacks an extended blue HB (Buzzoni et
al.~1992).  47 Tuc is known to have many types of exotic stars,
including one-third of all known millisecond pulsars (Robinson et
al.~1995), which are likely created by binary processes.  It is not
unreasonable that a small number of binary mergers could be
responsible for the supra-HB stars, given the cluster's large
total population, but we cannot test that possibility now.  Further optical
or UV photometry of the hot population will be needed to determine
whether a genuine blue HB is present.

We therefore believe that the hot population we have detected within
$2\, \rhalf$ in 47 Tuc consists (in declining luminosity order) of one
PAGB star (BS), 5--6 other stars in post-HB evolution, 9--15 hot HB
stars (somewhat fainter than expected from canonical theory), and a
few BSS objects at the threshold of our photometry.  This separation
is, however, not unambiguous, and colors will have to be obtained to
make positive classifications.  The small subsample with IUE or
optical data does tend to confirm this separation.

\section{Diffuse Far-UV Emission}

An unexpected feature of the UIT image is the presence of a diffuse
far-UV component in the cluster center (see Fig.~1).  The component is
also detectable on a shorter UIT exposure (FUV2388, exposure 1031 sec)
taken on a different orbit.  This may be the same component first
reported by Rich et al.~(1993) on IUE spectra of the inner 20\arcsec\
of the cluster, and our central surface brightness value agrees with
their IUE continuum value (but see qualifying comments below on the
IUE spectra).  The diffuse light is symmetrical about the cluster
center coordinates and is detectable up to a radius of 500\arcsec.
There is a low-level, asymmetric background, probably skyglow, on the
deeper UIT images which prevents us from making accurate measures at
larger radii; within 400\arcsec, however, the photometry is reliable.
The adopted background level is the mean flux in an annulus centered
on the cluster with inner and outer radii of 600 and 700\arcsec.
Tests show that instrumental scattered light from the bright PAGB star
has negligible influence on the cluster surface brightness at
distances larger than 25\arcsec\ from the star.  The integrated
brightness of the diffuse component is relatively large.  After all
resolved stars are masked out of the image, the diffuse light has 
$\mbfive = 10.2\pm 0.2$ within
\rhalf, corresponding to 40\% of the total FUV light in that region,
and $\mbfive = 9.4\pm 0.3$ within 400\arcsec\ (58\%).

The surface brightness profile of the diffuse component is shown in
Figure 5.  We have compared this distribution to that of the cluster's
integrated $V$-band light from Trager et al.~(1995) and computed the
\bfivev\ color profile also shown in Fig.~5.  There is a large color
gradient, from \bfivev\ values of $\sim 6$ mags at the center to
significantly bluer values $\sim 4$ mags by 300\arcsec\ radius.  This
means that the diffuse FUV light is significantly {\it less}
concentrated than the bulk of the stellar population.  Since the UV
background level is somewhat uncertain, we have checked that adopting
different background levels within a plausible range does not affect
the color gradient within a radius of 200\arcsec.

\begin{figure}
\plotone{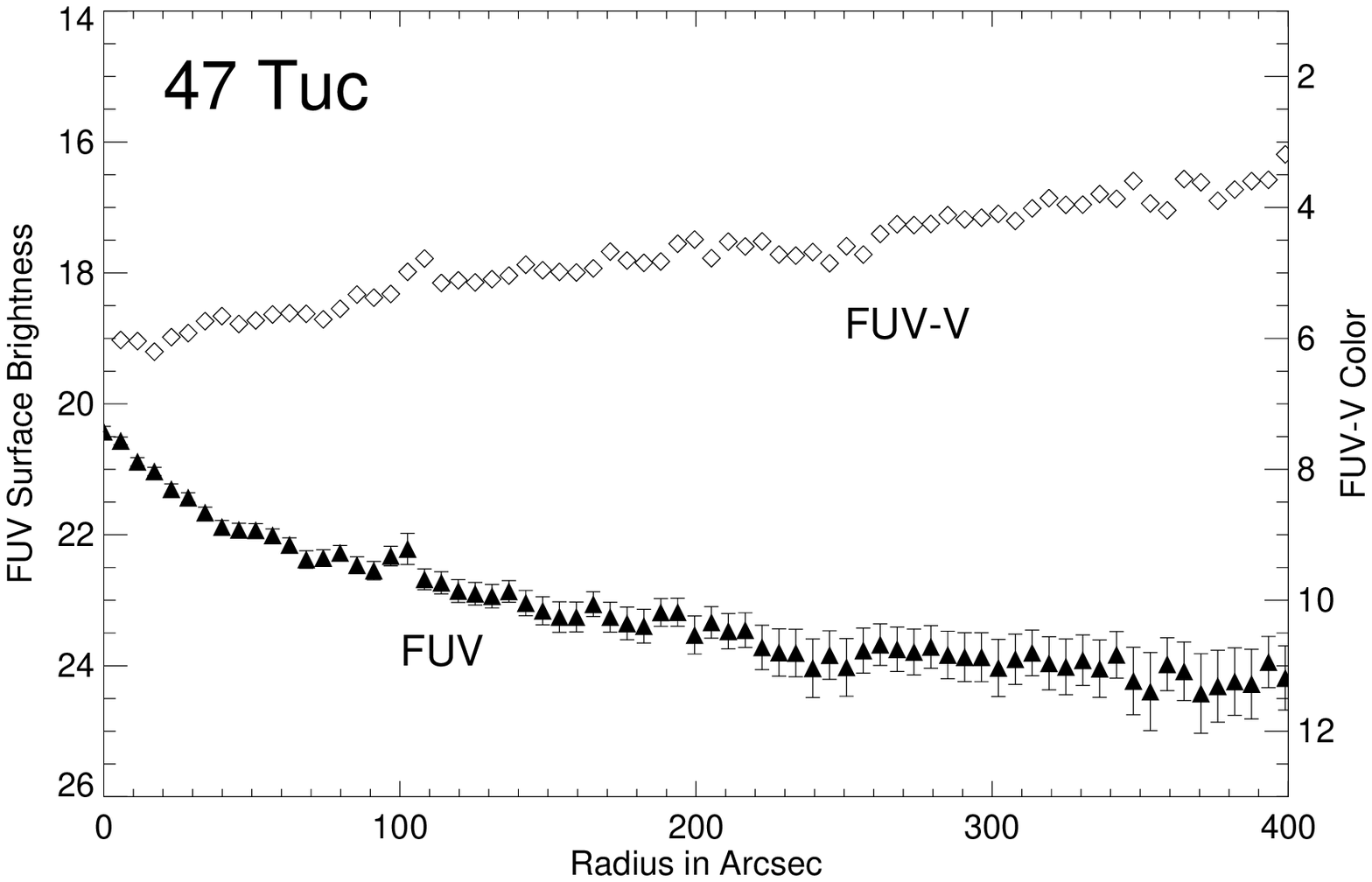}
\caption{ Surface photometry of the diffuse
far-UV component in 47 Tuc.  All stars have been masked out.  The
solid triangles show the mean surface brightness in circular annuli;
units (left hand scale) are monochromatic UV magnitudes per square
arcsecond.  The open diamonds are the FUV--V colors (right hand scale)
obtained by subtracting the Trager et al.\ (1995) V-band surface
brightnesses at the same locations.  Details on the background
correction are given in the text.} 
\end{figure}

Among clusters observed by UIT, only M79 has a similar diffuse feature
(Hill et al.~1996).  Most clusters on the UIT program have too many
UV-bright stars to detect a diffuse component easily. The central
regions of M79 are unresolved by UIT; beyond 1\arcmin\ from the core a
diffuse component is detectable at a surface brightness $\la 1$ mag
fainter than the total light.  Unlike the source in 47 Tuc, the
diffuse UV light of M79 appears to parallel the V-band light
distribution.

The simplest explanation for the diffuse UV component in 47 Tuc would be
that it is the combined light of the warm stars which are just below
the threshold for detection as individual objects in Fig.~4.  However,
in the case of normal warm HB stars, one would expect the FUV light to
follow the distribution of $V$-band light; and in the case of BSS
stars, it should be much more concentrated to the cluster center than
the $V$-band light (Paresce et al.~1991).  In contrast, the diffuse FUV
component is more extended than the $V$-band light.  
Some binary models for hot cluster stars produce such distributions
(e.g.~Bailyn et al.~1992; Bailyn 1995), but it seems unlikely that
there would be enough such objects just below the UIT threshold to
produce the observed diffuse source.

Another possibility is that the diffuse far-UV light is not stellar.
Krockenberger \& Grindlay (1995) have recently discovered a
bow-shock-like feature associated with 47 Tuc using ROSAT in the
0.1--0.4 keV X-ray band.  We do not detect any far-UV enhancement near
the X-ray feature (about $6\farcm 5$ NE of the cluster center), but
the presence of a possible bow shock suggests that 47 Tuc contains a
significant component of interstellar gas.  Presumably, this is the
remains of the stellar envelopes which have been shed during RGB mass
loss.  The most likely far-UV signature of this material would be the
C IV 1550 \AA\ emission feature.  Assigning all of the diffuse
emission to C IV, we have estimated the gas temperature and density
necessary to produce the far-UV surface brightness assuming that the
intracluster gas is collisionally ionized and in pressure equilibrium
with the shocked material at the interface with the Galactic halo.
For the carbon abundance of 47 Tuc, we find that an electron
temperature of $\sim 50000\,$K and $n_e \sim 0.05\, {\rm cm}^{-3}$
over \rhalf\ would yield the observed surface brightness.  The total
amount of material needed ($\sim 0.3\msun$) corresponds 
to the accumulation of mass lost from
the RGB (assuming 0.2\msun\ lost per giant) over a period of $\sim 3
\times 10^5$ years, roughly consistent with the residence time for gas
in the cluster given its velocity with respect to its surroundings.
Therefore, on the basis of its X-ray properties and the estimated RGB
mass loss rate, C IV emission from an intracluster medium is a
plausible source of the diffuse far-UV light in 47 Tuc.

Unfortunately, C IV emission is not detected on three large aperture
(10$\times$20\arcsec) spectra of the center of 47 Tuc in the IUE
archives.  Two of these (SWP 1510 and 2086) were discussed by Rich et
al.~(1993).  They show an extended but nonuniform far-UV continuum at
a level of about $5 \times 10^{-15}\,$\flun, which is consistent with
the mean far-UV surface brightness determined by UIT for the cluster
center ($2.3\times 10^{-17}\,\flun {\rm arcsec}^{-2}$) when integrated
over the IUE aperture.  No emission lines are visible.  These spectra
are from early in the IUE mission before operations procedures had
been standardized; they are noisy, and SWP 1510 is out of focus.  A
later spectrum, SWP 11126, with an exposure time of 49.5 kilosec,
shows no detectable far-UV flux at all, with an upper limit of
$3\times 10^{-15}\,$\flun.  These results are contradictory.  Ordinarily,
one would give more weight to the more recent spectrum, but it is
possible that the difficulty in centering IUE on a diffuse object led
to a pointing error in that case.  An offset of order 20\arcsec\ would
reduce the mean diffuse flux a factor of two.  Even in that case,
however, if the far-UV flux detected by UIT were concentrated in the C
IV emission doublet, the resulting feature would have been easily
detected.  Because of the inconsistencies in the IUE data, it is
premature to rule out the C IV interpretation, but the available spectra
do not support it.  

A final possibility is that the diffuse UV light is produced by
scattering of stellar UV photons by dust grains.  Gillett et al.
(1988) detected a small excess over photospheric emission at 100$\mu$
in 47 Tuc with IRAS, which they attributed to intracluster dust grains
heated by the cluster's integrated starlight.  They estimated the
total dust mass at $3\times 10^{-4}\msun$.  But they pointed out that
this is 100--1000 times smaller than expected given the standard RGB
injection rate and a residence time of $3\times 10^7$ yrs based on the
interval between Galactic plane crossings.  However, if the residence
time is reduced to only a few $10^5$ years by virtue of the
cluster-halo interaction demonstrated by Krockenberger \& Grindlay
(1995), then the dust mass is in much better agreement with
predictions.  In the UV, dust grains are strongly forward scattering
(Witt et al.~1992), and any UV-bright source on the far side of the
cluster center might well produce a detectable diffuse UV component.
The obvious candidate is the very bright PAGB star BS.  Whether
the brightness and symmetry of the diffuse light is consistent with
scattering cannot be decided without quantitative modeling.  One might
expect the grains produced in a sub-solar abundance system like 47 Tuc
to more resemble those in the SMC than those typical of
our Galaxy (e.g.\ Hutchings 1982, Mathis 1990).

\section{Summary and Discussion}

\subsection{Hot Stars}

UIT images at 1500 \AA\ of a 40\arcmin\ diameter field centered on the
metal rich globular cluster 47 Tuc have disclosed a population of 51
hot stars, many of which are probably cluster members.  We do not have
colors for most of these objects, so our analysis is based on their UV
luminosity function.  About 20 of the sample lie near or above the
predicted UV luminosity of the hot horizontal branch (HB) and are
within two half-light radii of the cluster center.  IUE spectra of two
of these are consistent with post-HB phases.  Overall, the sample
within $2\, \rhalf$ in 47 Tuc probably consists (in declining
luminosity order) of one PAGB star (BS), 5--6 other stars in post-HB
evolution, 9--15 hot HB stars (somewhat fainter than expected from
canonical theory), and a few BSS objects at the threshold of our
photometry.  This separation requires confirmation by multicolor
photometry.  We cannot rule out the possibility that the supra-HB
stars are merged binaries rather than normal post-HB objects.  

The UV-bright population of 47 Tuc demonstrates that populations with
metallicities as high as those of metal-rich globular clusters and
with dominant cool horizontal branches can nonetheless produce hot HB
stars.  Extreme HB stars and their subsequent hot He-shell-burning
evolutionary phases are the likely source of the ``UV-upturns'' seen
in elliptical galaxies (Greggio \& Renzini 1990; Dorman et al.~1995;
Brown, Ferguson, \& Davidsen~1996).  This type of object evidently
exists in 47 Tuc, even if the numbers are too small to produce a net
upturn in its far-UV energy distribution.

Over the past 10 years, deep optical and UV observations have shown
that many low metallicity globular clusters have bimodal or extended
horizontal branch mass distributions which include objects at high
temperature.  Examples include NGC 2808 (Ferraro et al.~1990; Sosin et
al.~1997), NGC 1851 (\cite{wa92}), M15 (\cite{cro88}), $\omega$ Cen
(Whitney et al.~1994), M79 (Hill et al.~1996), NGC 6752 (Landsman et
al.~1996), and NGC 362 (Dorman et al.~1997).  Our UV observations of
47 Tuc and recent optical-band HST imaging of the centers of two other
metal rich clusters NGC~6388 and NGC~6441 (Piotto et al.~1997, Rich et
al.~1997) show that this phenomenon extends to clusters of high
metallicity. Indeed, the observations of NGC~6388 and NGC~6441 (which
were not included in the UIT program because of heavy UV foreground
extinction) show that the fraction of hot HB stars in metal rich
clusters can be significant.
  
For the purposes of understanding the galaxies, it is important to
understand how such hot objects are produced in clusters, particularly
metal-rich clusters.  Is red giant branch mass loss in single stars
alone responsible, or are multi-object processes, such as binary mass
exchange, binary mergers, or dynamical interactions with neighbors
important (Iben 1990, Fusi-Pecci et al.~1993, Bailyn 1995)?  Do
several modes operate within a given cluster?  There is some evidence
that dynamical interactions promote hot HB's in some instances
(Fusi-Pecci et al.~1993).  On the other hand, the largest population
of EHB and post-EHB stars yet found resides in the low density cluster
$\omega$ Cen (Whitney et al.~1994).  Rich et al.~(1997) tend to favor
the dynamical explanation for the HB's of NGC 6388 and 6441, despite
the absence of the expected strong radial gradients in the blue
population.  47 Tuc is nearly as dense as these two objects and
contains hot stars, yet its hot population is much smaller.  Also, the
surface density of the UIT sample of hot stars in 47 Tuc, when
normalized to the Trager et al.~(1995) V-band surface brightness,
increases by at least a factor of two from the region inside \rhalf\
to regions at 1--3 \rhalf.  This is again contrary to expectations for
most multi-object processes.  

Whatever the clusters may be telling us about the possible dynamical
origin of hot HB stars, the dynamical environment of elliptical
galaxies is very different from clusters.  Striking variations in UV
properties are observed among the galaxies, in the form of differences
in the central UV colors (Burstein et al.~1988) and in strong
gradients in UV colors (O'Connell et al.~1992, Ohl et al.~1997).  It
seems unlikely that multi-object processes in the general
gravitational field of the galaxies are important to producing these
effects.  If the UV-bright stars in the galaxies were never members of
dense stellar aggregates, then the clusters most relevant to the
galaxies are those without strong dynamical effects on their hot star
populations.  On the other hand, if the hot stars originated in
globular clusters which were later destroyed by tidal interactions,
then the dynamical properties of the parent clusters could play a role
if a large proportion of close binaries were generated.  In
any event, until the origin of HB mass distributions is better
understood, through the study of the clusters, it is premature to try
to derive global properties such as age or abundance from the
UV-upturns of galaxies.

\subsection{Diffuse Light}

The UIT images also reveal a diffuse source in 47 Tuc which can be
traced to radii $\ga 500 \arcsec$.  The UV light is more extended than
the V-band light, which produces a strong UV--V color gradient, with
bluer colors at larger radii.  There is no good stellar candidate for
the source of this light.  We consider possible non-stellar radiation
from an intracluster medium which, from ROSAT and IRAS observations,
does seem to be present in 47 Tuc.  Line emission from C IV is
plausible based on expected physical conditions in the intracluster
medium.  The line is not confirmed by IUE spectra, but these have
inconsistent characteristics.  Scattering by dust grains of light from
the bright PAGB star is another possibility. 

\acknowledgments

We are grateful to Kent Montgomery and Kenneth Janes for providing
their CCD photometry tables to us in advance of publication, and to
Craig Sosin and George Djorgovski for pre-release HST data on 47 Tuc.
Bob Cornett and Bob Hill provided valuable technical information on
UIT performance.  We also thank Yoji Kondo for promptly granting
discretionary IUE time.  Parts of this research have been supported by
NASA grants NAG5-700 and NAGW-4106 to the University of Virginia.


\onecolumn

\begin{figure}
\setcounter{figure}{0}
\includegraphics{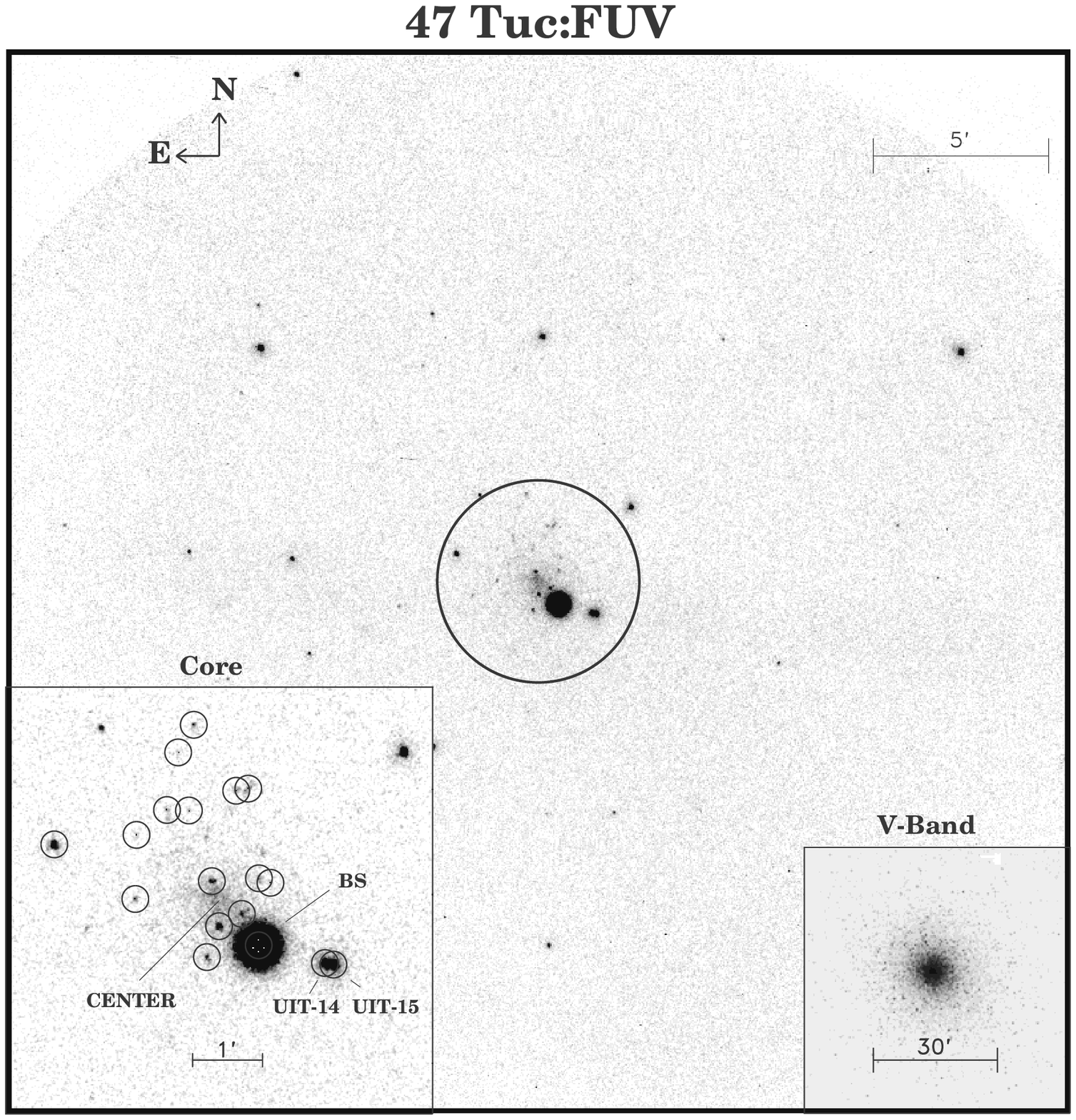}
\vspace{5.5truein}
\caption{ (Plate XX) The Ultraviolet Imaging
Telescope far-ultraviolet (1620 \AA) image of 47 Tuc.  The full field
of view is 40\arcmin\ in diameter; the edge of the field is visible at
the top.  The heavy line in the main image shows the circle containing
half the cluster light at optical wavelengths.  The inset at the lower
right shows the cluster in the V band; the inner few
arcminutes are burned out by cool main sequence and giant-branch stars
which are suppressed in the UV.  The inset at the left is an
enlargement of the center of the UV image showing all of the UV
identifications within the half-light radius.  The diffuse light
component discussed in \S 4 is just visible on the main and UV inset
images. }
\end{figure}

\end{document}